# Preliminary Evidence: Diagnosed Alzheimer's Disease But Not MCI Affects Working Memory Capacity: 0.7 of 2.7 Memory Slots is Lost

Eugen Tarnow, Ph.D.[1]

Avalon Business Systems, Inc.

18-11 Radburn Road, Fair Lawn, NJ 07410, USA

etarnow@avabiz.com

## Abstract

Recently it was shown explicitly that free recall consists of two stages: the first few recalls empty working memory (narrowly defined) and a second stage, a reactivation stage, concludes the recall (Tarnow, 2015; for a review of the theoretical predictions see Murdock, 1974). It was also shown that the serial position curve changes in mild Alzheimer's disease – lowered total recall and lessened primacy - are similar to second stage recall and different from recall from working memory.

The Tarnow Unchunkable Test (TUT, Tarnow, 2013) uses double integer items to separate out only the first stage, the emptying of working memory, by making it difficult to reactivate items due to the lack of intra-item relationships.

Here it is shown that subject TUT selects out diagnosed Alzheimer's Disease but not MCI. On average, diagnosed Alzheimer's Disease is correlated with a loss of 0.7 memory slots (out of an average of 2.7 slots).

The identification of a lost memory slot may have implications for improved stage definitions of Alzheimer's disease and for remediation therapy via working memory capacity management. In conjunction with the Alzheimer's disease process map, it may also be useful to identify the exact location of working memory.

---

[1] The author is an independent researcher. He received a PhD in physics from MIT. His research areas include short term memory, conformity, obedience, dreams, marketing science, management science, and semiconductor physics.

"Preliminary Evidence: Diagnosed Alzheimer's Disease But Not MCI affects the Capacity of Working Memory: 0.7 of 2.7 Memory Slots Is Lost" by Eugen Tarnow

"Preliminary Evidence: Diagnosed Alzheimer's Disease But Not MCI affects the Capacity of Working Memory: 0.7 of 2.7 Memory Slots Is Lost" by Eugen Tarnow

# Introduction

Free recall, in which items in a list are displayed or read to subjects who are then asked to retrieve the items, is one of the simplest ways to probe short term memory. It is used in neuropsychological batteries to test for the presence of Alzheimer's disease, for example, MMSE, ADAS-Cog, FREES, CVLT, DWR, etc (for a review see Cullen et al, 2007). The corresponding serial position curve, the probability of recalling an item versus the order in which the item was presented, is u-shaped: items in the beginning of the presented list (primacy) and at the end of the list (recency) are more likely to be recalled than those in the middle of the list.

It was recently shown explicitly that free recall is a well defined two stage process (Tarnow, 2015; this had been suggested before, for a review see Murdock, 1974). In the first stage working memory is emptied and in the second stage a different retrieval process occurs using reactivation of items. Working memory is responsible for recency (and some primacy for short lists) and the second stage recall shows some primacy but no recency. Bayley et al (2000) found that a reduction in the primacy effect is an early and ubiquitous feature of the memory impairment of AD and Tarnow (2016) showed that early Alzheimer's disease correlates with a loss of items from the second stage. In this contribution we explore what happens to the first stage in MCI and diagnosed Alzheimer's disease.

# Method

### Subjects

Participants (see Table 1) were part of a clinical study for aging and dementia at the Alzheimer's Disease Research Center at Icahn School of Medicine at Mount Sinai (ADRC ISMMS). Inclusion criteria included 65 years of age or older, primarily English/Chinese/Spanish speaking, visual and auditory acuity adequate for cognitive testing, willingness to participate in all clinical assessment, and having a study partner available as an informant.

### Study procedure

All participants were asked to complete a 3-hour in-person dementia evaluation, which consisted of recording of demographic information, clinical interview, cognitive evaluation, medical examination, and functional assessment. Upon completion of the evaluation, all participants were assigned a research diagnosis of normal cognition, Mild Cognitive Impairment (MCI), and dementia using Clinical Research

"Preliminary Evidence: Diagnosed Alzheimer's Disease But Not MCI affects the Capacity of Working Memory: 0.7 of 2.7 Memory Slots Is Lost" by Eugen Tarnow

Diagnostic Criteria employed at the ADRC ISMMS.  The study was approved by the institutional board of the ISMMS.  All participants provided signed informed consent.

Records indicating dementia due to oncology treatment and due to vascular dementia were discarded as was a record of a subject who refused to answer more than once.

**DX**

|  |  | Frequency | Percent | Valid Percent | Cumulative Percent |
|---|---|---|---|---|---|
| Valid | Normal | 83 | 62.9 | 63.8 | 63.8 |
|  | MCI | 37 | 28.0 | 28.5 | 92.3 |
|  | AD | 10 | 7.6 | 7.7 | 100.0 |
|  | Total | 130 | 98.5 | 100.0 |  |
| Missing | System | 2 | 1.5 |  |  |
| Total |  | 132 | 100.0 |  |  |

*Table 1.  Breakdown of study subjects.*

The Tarnow Unchunkable Test (TUT, Tarnow, 2013) attempts to separate out just the first stage of free recall which empties working memory.  It uses particular double-digit combinations which lack intra-item relationships, minimizing inter-item associative strengths (Deese, 1959), so that the second reactivation stage does not occur.   The TUT is copyrighted and patent pending and can be purchased from Tarnow. It consists of 3-item tests and 4-item tests and can be done using paper, as was the case in this article, or computer.  It has been used on 193 Russian college students (Ershova & Tarnow, 2016). The time to administer the 3 item test by itself is a couple of minutes.

"Preliminary Evidence: Diagnosed Alzheimer's Disease But Not MCI affects the Capacity of Working Memory: 0.7 of 2.7 Memory Slots Is Lost" by Eugen Tarnow

# Results

The main results will center on the 3 item memory score.

The properties of the different cultures of the normal subjects are shown in Table 2. The 3 and 4 item memory tests are culture independent. In the subject sample age and gender are rather evenly distributed among the cultures. The only cultural dependency we find is the educational level which is highest for English and Mandarin and lower for Cantonese and then for Spanish.

Grouping field: Language
*Cells contain: Mean, Standard Deviation, Standard Error, Count

| Field | English* | Cantonese* | Mandarin* | Spanish* | F-Test | df | Importance |
|---|---|---|---|---|---|---|---|
| Education (total years) | 15.971 | 12.158 | 15.000 | 7.500 | 8.992 | 3, 80 | 1.000 Important |
|  | 3.316 | 4.207 | 2.171 | 4.950 |  |  |  |
|  | 0.569 | 0.965 | 0.403 | 3.500 |  |  |  |
|  | 34 | 19 | 29 | 2 |  |  |  |
| Age | 76.029 | 71.947 | 73.345 | 76.500 | 2.047 | 3, 80 | 0.886 Unimportant |
|  | 6.987 | 6.160 | 5.459 | 4.950 |  |  |  |
|  | 1.198 | 1.413 | 1.014 | 3.500 |  |  |  |
|  | 34 | 19 | 29 | 2 |  |  |  |
| Gender | 1.824 | 1.737 | 1.655 | 2.000 | 1.007 | 3, 80 | 0.606 Unimportant |
|  | 0.387 | 0.452 | 0.484 | 0.000 |  |  |  |
|  | 0.066 | 0.104 | 0.090 | 0.000 |  |  |  |
|  | 34 | 19 | 29 | 2 |  |  |  |
| TUT: 4 item memory | 2.123 | 1.907 | 2.069 | 1.500 | 0.581 | 3, 79 | 0.371 Unimportant |
|  | 0.749 | 0.920 | 0.773 | 1.650 |  |  |  |
|  | 0.128 | 0.217 | 0.144 | 1.167 |  |  |  |
|  | 34 | 18 | 29 | 2 |  |  |  |
| TUT: 3 item memory | 2.637 | 2.509 | 2.552 | 2.667 | 0.227 | 3, 80 | 0.123 Unimportant |
|  | 0.540 | 0.602 | 0.668 | 0.471 |  |  |  |
|  | 0.093 | 0.138 | 0.124 | 0.333 |  |  |  |
|  | 34 | 19 | 29 | 2 |  |  |  |

*Table 2. ANOVA of the language groups show language independence of the TUT 3 and 4 item memory tests for normal subjects. Educational level was significantly dependent upon the language.*

"Preliminary Evidence: Diagnosed Alzheimer's Disease But Not MCI affects the Capacity of Working Memory: 0.7 of 2.7 Memory Slots Is Lost" by Eugen Tarnow

In Table 3 is shown that the 3 item memory and 4 item memory tests are not correlated with gender.  In Table 4 is shown that the TUT tests do not correlate with age or education.  In addition the 3 and 4 item tests are not strongly correlated.  It was found elsewhere (Ershova & Tarnow, 2016) that it is because the 4 item test exceeds the capacity of most subjects' working memory and the subjects do not manage their working memory efficiently under those circumstances with a resulting large distribution of scores.

Grouping field: Gender
*Cells contain: Mean, Standard Deviation, Standard Error, Count

| Field | Male* | Female* | F-Test | df | Importance |
|---|---|---|---|---|---|
| TUT: 3 item ... | 2.635 | 2.581 | 0.136 | 1, 81 | 0.287 Unimportant |
| | 0.682 | 0.546 | | | |
| | 0.149 | 0.069 | | | |
| | 21 | 62 | | | |
| TUT: 4 item ... | 2.117 | 2.051 | 0.106 | 1, 80 | 0.254 Unimportant |
| | 0.767 | 0.790 | | | |
| | 0.171 | 0.100 | | | |
| | 20 | 62 | | | |

*Table 3.  ANOVA test for Gender.*

| *Adjusted R squared* | TUT: 3 item | TUT: 4 item |
|---|---|---|
| Age | -0.006 | -0.012 |
| Education | 0.033 | 0.062 |
| TUT: 3 item | | .168 |
| TUT: 4 item | | |

*Table 4. Regression shows the test results for normal subjects are independent of age and education and within the test only the TUT: 3 and 4-item tests are somewhat correlated.*

In contrast, statistical tests on the AD population versus the others (normal and MCI) (see Table 5) show that the 3 item memory is strongly correlated with the AD diagnosis; age is as well. The AD population average is 0.7 items below the normal population (Table 6). Statistical tests on the MCI population (Table 7) shows that neither the 3 item memory or the 4 item memory are correlated with the MCI diagnosis.

Grouping field: AD
*Cells contain: Mean, Standard Deviation, Standard Error, Count

| Field | Other* | AD* | F-Test | df | Importance |
|---|---|---|---|---|---|
| TUT: 3 item memory | 2.550<br>0.568<br>0.052<br>120 | 1.933<br>0.625<br>0.198<br>10 | 10.729 | 1, 128 | 0.999<br>⭐ Important |
| Age | 74.825<br>6.206<br>0.567<br>120 | 81.400<br>8.154<br>2.579<br>10 | 9.857 | 1, 128 | 0.998<br>⭐ Important |
| Gender | 1.697<br>0.461<br>0.042<br>119 | 1.500<br>0.527<br>0.167<br>10 | 1.655 | 1, 127 | 0.799<br>▫ Unimportant |
| TUT: 4 item memory | 2.015<br>0.806<br>0.074<br>119 | 1.667<br>1.077<br>0.341<br>10 | 1.636 | 1, 127 | 0.797<br>▫ Unimportant |
| Education (total years) | 14.500<br>7.190<br>0.656<br>120 | 11.800<br>5.308<br>1.679<br>10 | 1.345 | 1, 128 | 0.752<br>▫ Unimportant |

*Table 5. ANOVA distinguishing AD diagnosed subjects from remaining subjects.*

"Preliminary Evidence: Diagnosed Alzheimer's Disease But Not MCI affects the Capacity of Working Memory: 0.7 of 2.7 Memory Slots Is Lost" by Eugen Tarnow

Grouping field: Dx
*Cells contain: Mean, Standard Deviation, Standard Error, Count

| Field | Normal* | MCI* | AD* | F-Test | df | Importance |
|---|---|---|---|---|---|---|
| TUT: 3 item ... | 2.594 | 2.450 | 1.933 | 6.205 | 2, 127 | 0.997 |
|  | 0.580 | 0.534 | 0.625 |  |  | ★ Important |
|  | 0.064 | 0.088 | 0.198 |  |  |  |
|  | 83 | 37 | 10 |  |  |  |

*Table 6. Average scores on 3 item test for the different diagnostic categories. The difference between AD and Normals is 0.7 items.*

Grouping field: MCI
*Cells contain: Mean, Standard Deviation, Standard Error, Count

| Field | Other* | MCI* | F-Test | df | Importance |
|---|---|---|---|---|---|
| Language | 2.054 | 2.351 | 2.699 | 1, 128 | 0.897 |
|  | 0.971 | 0.824 |  |  | • Unimportant |
|  | 0.101 | 0.135 |  |  |  |
|  | 93 | 37 |  |  |  |
| Gender | 1.720 | 1.583 | 2.254 | 1, 127 | 0.864 |
|  | 0.451 | 0.500 |  |  | • Unimportant |
|  | 0.047 | 0.083 |  |  |  |
|  | 93 | 36 |  |  |  |
| Age | 75.032 | 76.081 | 0.671 | 1, 128 | 0.586 |
|  | 6.923 | 5.634 |  |  | • Unimportant |
|  | 0.718 | 0.926 |  |  |  |
|  | 93 | 37 |  |  |  |
| TUT: 4 item memory | 2.024 | 1.901 | 0.574 | 1, 127 | 0.550 |
|  | 0.820 | 0.860 |  |  | • Unimportant |
|  | 0.085 | 0.141 |  |  |  |
|  | 92 | 37 |  |  |  |
| TUT: 3 item memory | 2.523 | 2.450 | 0.397 | 1, 128 | 0.470 |
|  | 0.617 | 0.534 |  |  | • Unimportant |
|  | 0.064 | 0.088 |  |  |  |
|  | 93 | 37 |  |  |  |
| Education (total years) | 14.301 | 14.270 | 0.000 | 1, 128 | 0.018 |
|  | 3.953 | 11.827 |  |  | • Unimportant |
|  | 0.410 | 1.944 |  |  |  |
|  | 93 | 37 |  |  |  |

"Preliminary Evidence: Diagnosed Alzheimer's Disease But Not MCI affects the Capacity of Working Memory: 0.7 of 2.7 Memory Slots Is Lost" by Eugen Tarnow

*Table 7. ANOVA distinguishing MCI from remaining subjects.*

The serial position curves for the normal and AD populations for the 3 item test are shown in Fig. 1. They show primacy and no recency (also in Ershova & Tarnow, 2016). The first item is not very affected by AD but the second and third items are much less likely to be recalled.

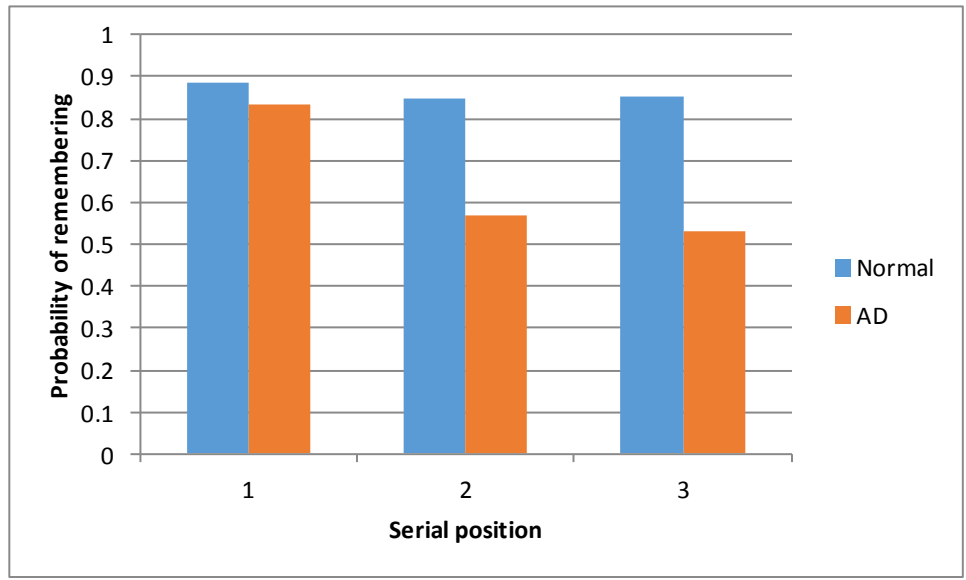

*Fig. 1. Serial position curves for 3 item test.*

Total recall distributions for the 3 item tests is shown in Fig. 2. The distribution for MCI is similar to normal while the distribution for AD subjects is moved towards lower values. The corresponding ROC curve is shown in Fig. 4 and the area under the curve is 0.78. Fig. 5 shows how much the 3 item score increases the odds of an AD diagnosis in the current subject population.

Total recall distributions for the 4 item tests is shown in Fig. 6. The distributions are much wider (also in Ershova & Tarnow, 2016) which presumably explains why it is not sensitive to AD even though the serial position curves show similar differences between AD and normal subjects (Fig. 7).

"Preliminary Evidence: Diagnosed Alzheimer's Disease But Not MCI affects the Capacity of Working Memory: 0.7 of 2.7 Memory Slots Is Lost" by Eugen Tarnow

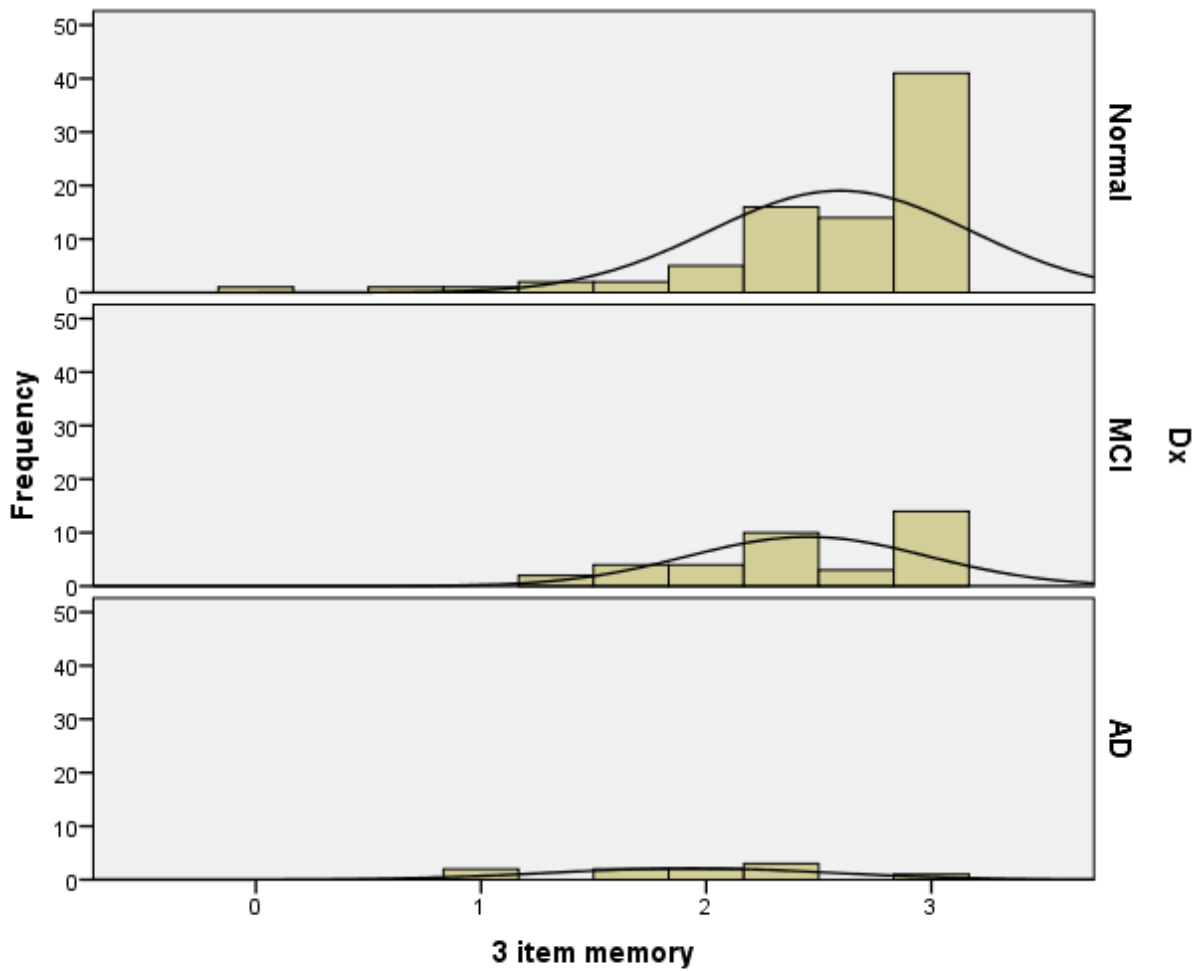

*Figure 2. Histograms for 3 item memory.*

"Preliminary Evidence: Diagnosed Alzheimer's Disease But Not MCI affects the Capacity of Working Memory: 0.7 of 2.7 Memory Slots Is Lost" by Eugen Tarnow

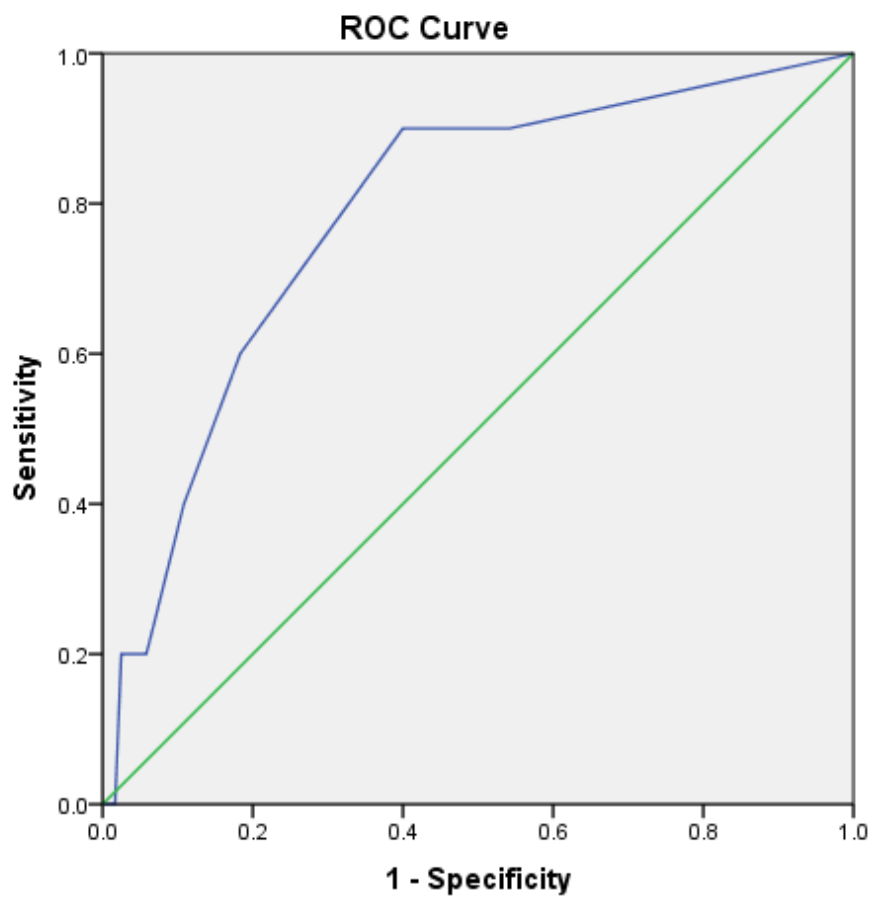

*Figure 3. ROC area under curve for AD diagnostic stage=0.78.*

"Preliminary Evidence: Diagnosed Alzheimer's Disease But Not MCI affects the Capacity of Working Memory: 0.7 of 2.7 Memory Slots Is Lost" by Eugen Tarnow

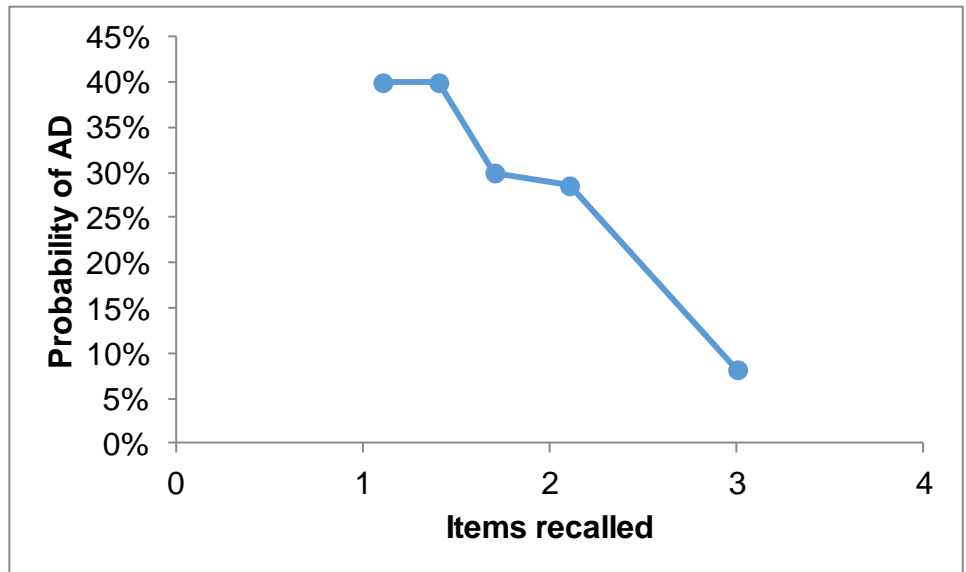

*Fig. 4. Probability that a subject coming to the clinic will have AD as a function of the TUT 3 item recall.*

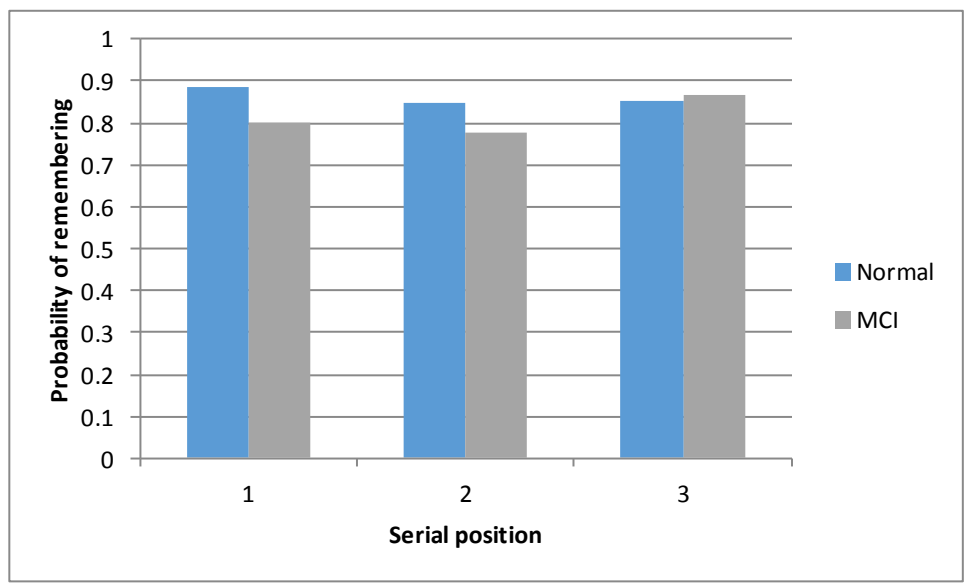

*Fig. 5. Serial position curves for 3 item test for normal and MCI subjects.*

"Preliminary Evidence: Diagnosed Alzheimer's Disease But Not MCI affects the Capacity of Working Memory: 0.7 of 2.7 Memory Slots Is Lost" by Eugen Tarnow

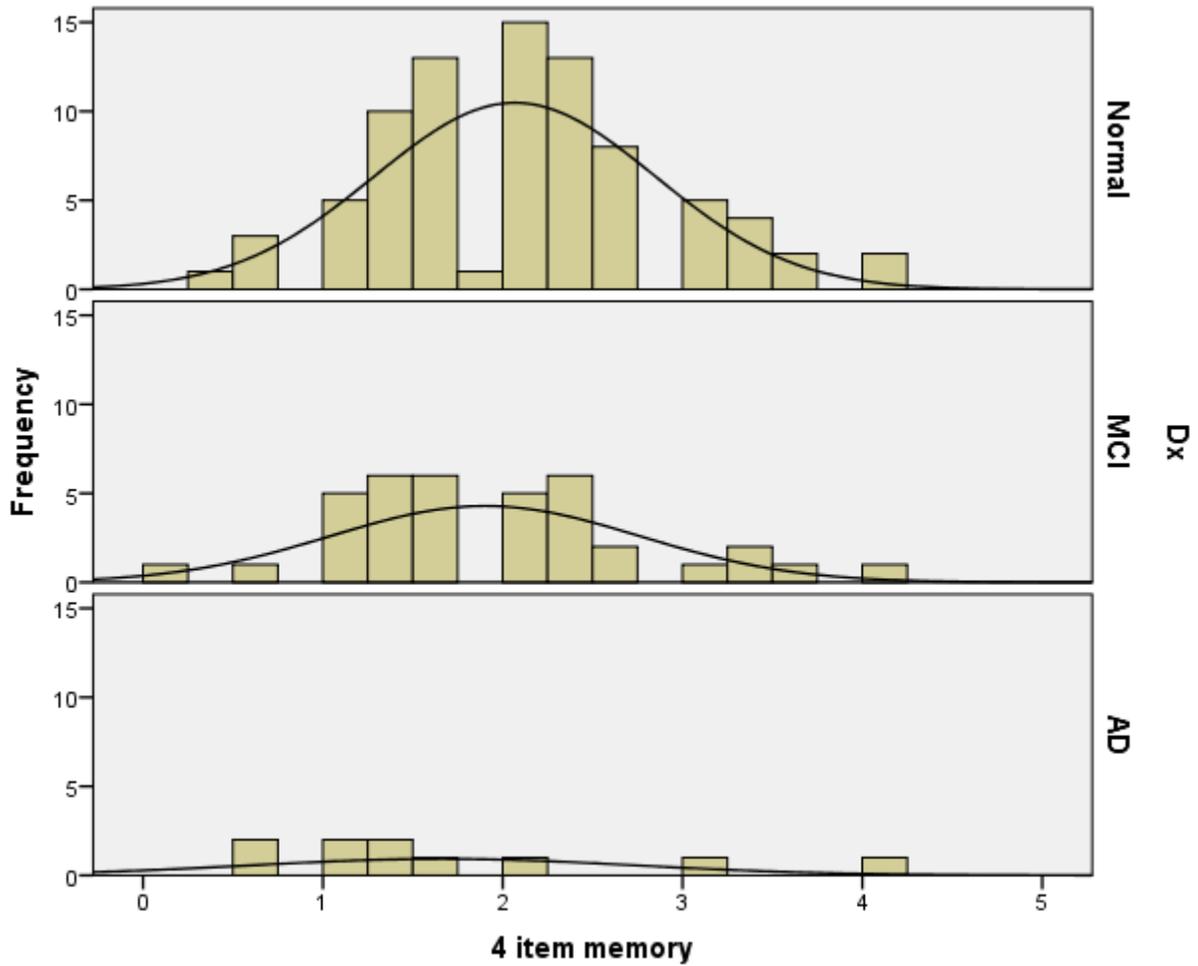

*Fig 6. Histograms of TUT 4-item memory recalls.*

"Preliminary Evidence: Diagnosed Alzheimer's Disease But Not MCI affects the Capacity of Working Memory: 0.7 of 2.7 Memory Slots Is Lost" by Eugen Tarnow

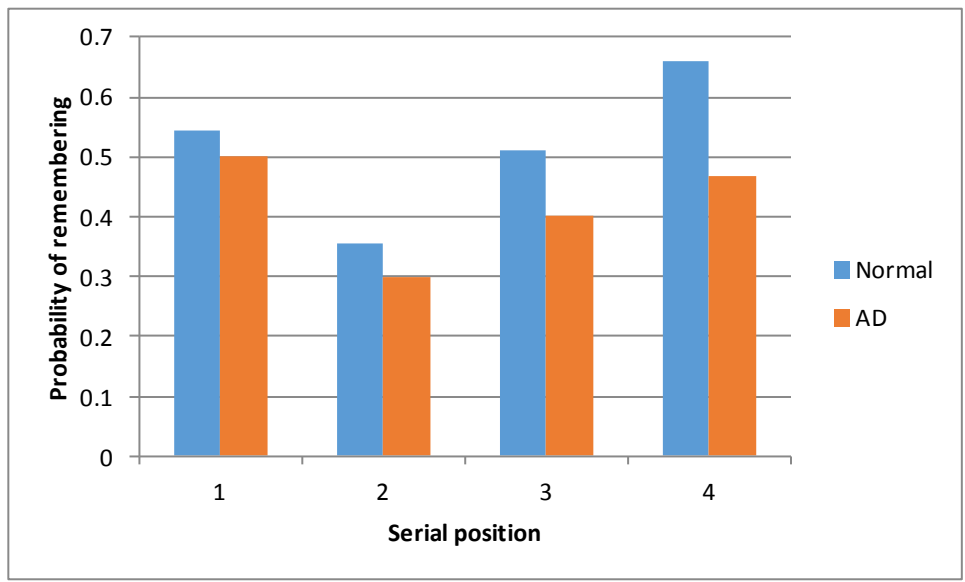

*Fig. 7. Serial position curves of normal and AD subjects for the TUT 4-item test.*

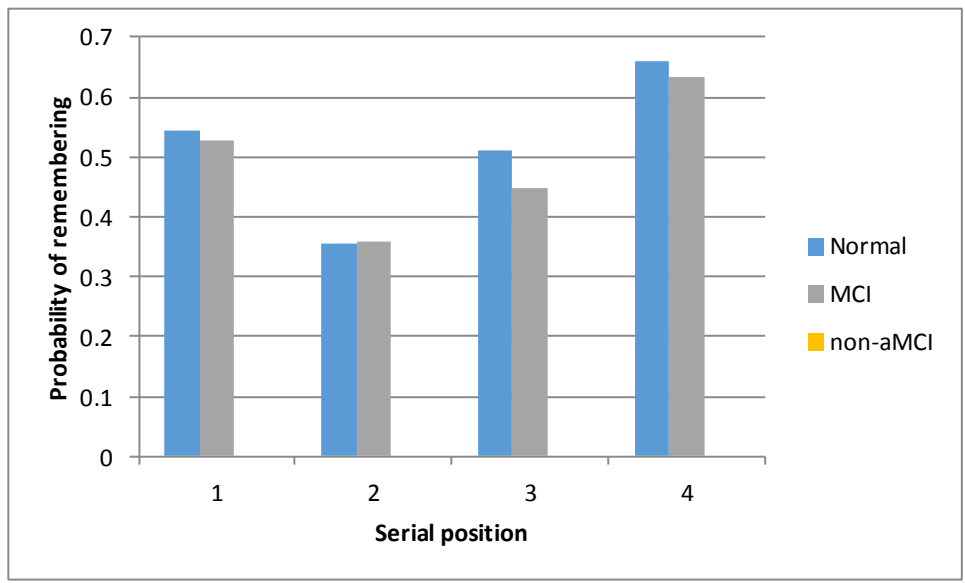

*Fig. 8. Serial position curves of normal and MCI subjects for the TUT 4-item test.*

## Discussion

The results show that the first stage of free recall as measured by the TUT 3 item test is not affected by MCI but by an AD diagnosis.  Damage to the second stage of free recall occurs in early AD (Tarnow, 2015) and damage to the first stage of recall occurs in diagnosed AD.

The finding that AD results in the loss of working memory may be important for remediation therapy, though working memory capacity management has yet to be tried on a large population.  But perhaps now is the time (this was argued to be important for a population of college students as well in Ershova & Tarnow, 2016).

The size of working memory may also be a measure of drug efficacy: if AD working memory changes can be prevented or reversed, then the drug may prevent or reverse the AD stage.

It could also be that the loss of working memory may be a good diagnostic milestone since it is well defined.  It presumably corresponds to a stage in the Braak & Braak (1995) observations.

That the TUT displays language independency would suggest that it is a good test to use on multicultural populations.  It also displays independency of age and education, further properties that are important for testing populations for physical diseases like Alzheimer's disease that may be in part hidden by aging or educational attainments.

"Preliminary Evidence: Diagnosed Alzheimer's Disease But Not MCI affects the Capacity of Working Memory: 0.7 of 2.7 Memory Slots Is Lost" by Eugen Tarnow

# Acknowledgement

The testing of the subjects was supported by the AG005183 (PI: Sano).

"Preliminary Evidence: Diagnosed Alzheimer's Disease But Not MCI affects the Capacity of Working Memory: 0.7 of 2.7 Memory Slots Is Lost" by Eugen Tarnow

"Preliminary Evidence: Diagnosed Alzheimer's Disease But Not MCI affects the Capacity of Working Memory: 0.7 of 2.7 Memory Slots Is Lost" by Eugen Tarnow